\documentclass[pra,aps,twocolumn,floatfix,showpacs]{revtex4}
\usepackage{graphics}
\begin{document}
\title{Population imbalanced fermions in harmonically trapped optical lattices}
\author{M. Iskin and C. J. Williams}
\affiliation{Joint Quantum Institute, National Institute of Standards and Technology, and 
University of Maryland, Gaithersburg, Maryland 20899-8423, USA.}
\date{\today}

\begin{abstract}

The attractive Fermi-Hubbard Hamiltonian is solved via the Bogoliubov-de Gennes 
formalism to analyze the ground state phases of population imbalanced fermion 
mixtures in harmonically trapped two-dimensional optical lattices. In the low 
density limit the superfluid order parameter modulates in the radial direction 
towards the trap edges to accommodate the unpaired fermions that are pushed 
away from the trap center with a single peak in their density. However in 
the high density limit while the order parameter modulates in the radial 
direction towards the trap center for low imbalance, it also modulates 
towards the trap edges with increasing imbalance until the 
superfluid to normal phase transition occurs beyond a critical imbalance. 
This leads to a single peak in the density of unpaired fermions for low 
and high imbalance but leads to double peaks for intermediate imbalance.

\end{abstract}
\pacs{03.75.Hh, 03.75.Kk, 03.75.Ss}
\maketitle

The phase diagram of dilute population imbalanced fermion mixtures
has been recently studied showing superfluid to normal phase transition 
with increasing imbalance as well as a phase separation between paired and 
unpaired fermions~\cite{mit, rice, mit-2, rice-2}. These recent works are 
extentions of the earlier works on dilute population balanced mixtures where 
a crossover from Bardeen-Cooper-Schrieffer (BCS) to Bose-Einstein 
condensation (BEC) type superfluidity is observed as a function of the 
attractive fermion-fermion interaction strength~\cite{djin, rgrimm, jthomas, csalomon, rhulet}.

Arguably understanding the phase diagram of fermion mixtures in optical 
lattices is one of the next frontiers in cold atoms research because of their 
great tunability. In addition to the particle populations and the particle-particle 
interaction strengths, one can also precisely control the particle 
tunnelings, the lattice dimensionality and the lattice geometry. For instance experimental 
evidence for the superfluid and the insulating phases of population balanced mixtures have 
been recently reported in trapped optical lattices~\cite{mit-lattice},
after overcoming some earlier difficulties~\cite{modugno, kohl, stoferle, bongs}.
This recent work has also opened the possibility of studying many-body 
properties of population imbalanced mixtures in optical lattices.

Earlier theoretical works on population imbalanced fermion mixtures in optical 
lattices were limited to homogenous systems~\cite{iskin-lattice, torma-lattice}, 
and they showed rich phase diagrams involving BCS type nonmodulating and 
Fulde-Ferrell-Larkin-Ovchinnikov (FFLO) type spatially modulating superfluid phases
in addition to insulating and normal phases. Furthermore the phase diagram of 
population imbalanced mixtures in harmonically trapped optical lattices has been
recently discussed within the semi-classical local density approximation 
(LDA)~\cite{koponen}. However it has been previously shown that the LDA type 
methods are not sufficient to describe even the dilute population imbalanced 
mixtures without an optical lattice~\cite{mizushima, torma}. 
In this manuscript we therefore analyze the ground state phases of fermion mixtures 
in harmonically trapped two-dimensional optical lattices via using the fully quantum 
mechanical Bogoliubov-de Gennes (BdG) method where the trapping potential
is included exactly at the mean-field level.

Our main results are as follows.
In the low density limit the superfluid order parameter modulates in the radial 
direction towards the trap edges to accommodate the unpaired fermions that 
are pushed away from the trap center with a single peak in their density. 
These findings are in good agreement with the recent 
theoretical~\cite{mizushima, torma, liu-mixture} and 
experimental~\cite{mit, rice, mit-2, rice-2} findings on dilute population 
imbalanced mixtures without an optical lattice. However in the high density limit 
while the order parameter modulates in the radial direction towards the trap 
center for low imbalance, it also modulates towards the trap edges with increasing 
imbalance until the superfluid to normal phase transition occurs beyond a 
critical imbalance. This leads to a single peak in the density of unpaired fermions 
for low and high imbalance but leads to double peaks for intermediate imbalance.

\textit{BdG formalism}:
To achieve these results we solve the Fermi-Hubbard Hamiltonian
\begin{eqnarray}
\label{eqn:hamiltonian}
H_{FH} = &-& \sum_{i,j,\sigma} t_{i,j,\sigma} a_{i,\sigma}^\dagger a_{j,\sigma}
-\sum_{i,\sigma} (\mu_\sigma - V_{i,\sigma}) a_{i,\sigma}^\dagger a_{i,\sigma} \nonumber \\
&-& \sum_{i,j} U_{i,j} a_{j,\uparrow}^\dagger a_{j,\uparrow} a_{i,\downarrow}^\dagger a_{i,\downarrow},
\end{eqnarray}
where $a_{i,\sigma}^\dagger$ ($a_{i,\sigma}$) creates (annihilates) a pseudo-spin 
$\sigma$ fermion at lattice site $i$, $t_{i,j,\sigma}$ and $U_{i,j} \ge 0$ are the 
particle-particle tunneling and the density-density interaction matrix elements, 
$\mu_\sigma$ is the chemical potential, and 
$V_{i,\sigma} = \alpha_\sigma |\mathbf{r_i}|^2/2$ is the trapping potential at 
position $\mathbf{r}_i$ with $\alpha_\sigma = m_\sigma \omega_\sigma^2$ 
such that the trapping potential is centered at the origin. Here the label $\sigma$ 
identifies $\uparrow$ or $\downarrow$ fermions and allows $\sigma$ fermions 
to have equal or unequal masses controlled by $t_{i,j,\sigma}$ and/or to have 
equal or unequal populations controlled by $\mu_\sigma$.

In the mean-field approximation for the superfluid phase, the Fermi-Hubbard 
Hamiltonian reduces to
$
H = - \sum_{i,j,\sigma} t_{i,j,\sigma} a_{i,\sigma}^\dagger a_{j,\sigma}
-\sum_{i,\sigma} (\mu_\sigma - V_{i,\sigma}) a_{i,\sigma}^\dagger a_{i,\sigma}
- \sum_{i,j} ( \Delta_{i,j} a_{j,\downarrow}^\dagger a_{i,\uparrow}^\dagger 
+ \Delta_{i,j}^* a_{i,\uparrow} a_{j,\downarrow} - |\Delta_{i,j}|^2 / U_{i,j} ),
$
where the self-consistent field
$
\Delta_{i,j} = U_{i,j} \langle a_{i,\uparrow} a_{j,\downarrow} \rangle 
$
is the superfluid order parameter and $\langle ... \rangle$ is a thermal average.
The mean-field Hamiltonian can be diagonalized via the Bogoliubov-Valatin transformation
$
a_{i,\sigma} = \sum_n [u_{n,i,\sigma} \gamma_{n,\sigma} - s_\sigma v_{n,i,\sigma}^* \gamma_{n,-\sigma}^\dagger],
$
where $\gamma_{n,\sigma}^\dagger$ ($\gamma_{n,\sigma}$) creates (annihilates) 
a pseudo-spin $\sigma$ quasiparticle with the wavefunction $u_{n,i,\sigma}$ 
($v_{n,i,\sigma}$), and $s_\uparrow = +1$ and $s_\downarrow = -1$. 
This leads to the BdG equations
\begin{equation}
\label{eqn:bdg}
\sum_j 
\left( \begin{array}{cc} 
T_{i,j,\uparrow} & \Delta_{i,j} \\ \Delta_{i,j}^* & -T_{i,j,\downarrow}^*
\end{array} \right)
\mathbf{\varphi_{n,j,\sigma}} = s_\sigma \epsilon_{n,\sigma} \mathbf{\varphi_{n,i,\sigma}},
\end{equation}
where
$
T_{i,j,\sigma} = -t_{i,j,\sigma} - (\mu_\sigma - V_{i,\sigma})\delta_{i,j}
$
is the diagonal element and $\delta_{i,j}$ is the Kronecker delta. Here 
$\epsilon_{n,\sigma} > 0$ are the eigenvalues and $\mathbf{\varphi_{n,i,\sigma}}$ 
are the eigenfunctions given by
$
\mathbf{\varphi_{n,i,\uparrow}}^\dagger = (u_{n,i,\uparrow}^*, v_{n,i,\downarrow}^*)
$
for the $\uparrow$ and
$
\mathbf{\varphi_{n,i,\downarrow}}^\dagger = (v_{n,i,\uparrow}, -u_{n,i,\downarrow})
$
for the $\downarrow$ eigenvalues. Since solutions to the BdG equations are 
invariant under the transformation
$v_{n,i,\uparrow} \to u_{n,i,\uparrow}^*$, $u_{n,i,\downarrow} \to -v_{n,i,\downarrow}^*$ 
and $\epsilon_{n,\downarrow} \to -\epsilon_{n,\uparrow}$, it is sufficient to 
solve only for $u_{n,i} \equiv u_{n,i,\uparrow}$, $v_{n,i} \equiv v_{n,i,\downarrow}$ 
and $\epsilon_n \equiv \epsilon_{n,\uparrow}$ as long as we keep all the solutions 
with positive and negative $\epsilon_n$.

In Eq.~(\ref{eqn:bdg}) the superfluid order parameter $\Delta_{i,j}$ is given by
$
\Delta_{i,j} = - \sum_{n} U_{i,j} u_{n,i} v_{n,j}^* f(\epsilon_n)
$
where $f(x) = 1/[\exp(x/T) + 1]$ is the Fermi function and $T$ is the temperature.
Notice that this equation is free from the ultraviolet divergence and therefore it
does not explicitly involve any energy cut-off since the lattice spacing provides 
an implicit cut-off. Eq.~(\ref{eqn:bdg}) and the order parameter equation have to 
be solved self-consistently with the number equations 
$
0 \le n_{i, \sigma} \le 1 = \langle a_{i,\sigma}^\dagger a_{i,\sigma} \rangle
$
for the $\sigma$ fermions, where
$
n_{i,\uparrow} = \sum_{n} |u_{n,i}|^2 f(\epsilon_n)
$
and
$
n_{i,\downarrow} = \sum_{n} |v_{n,i}|^2 f(-\epsilon_n)
$
such that $N_\sigma = \sum_i n_{i,\sigma}$ determines $\mu_\sigma$. In the following
we consider only attractive and onsite s-wave interactions and set $U_{i,j} = U_0 \delta_{i,j}$
with $U_0 \ge 0$. Notice that this leads to $\Delta_{i,j} = \Delta_i \delta_{i,j}$. 
Furthermore fermions are allowed to tunnel only to the nearest neighbor sites 
and thus $t_{i,j,\sigma} = t_\sigma \delta_{i,j\pm1}$.

\textit{Ground state phases}:
We now analyze the ground state phases of fermion mixtures with equal masses 
($m_0 = m_\uparrow = m_\downarrow$), equal tunnelings ($t_0 = t_\uparrow = t_\downarrow$) 
and equal trapping potentials ($\alpha_0 = \alpha_\uparrow = \alpha_\downarrow$) but with 
unequal chemical potentials. The theoretical parameters $t_0$ and $U_0$ can be expressed in 
terms of the experimental parameters of the two-dimensional optical lattice potential 
$
V_L(x, y) = V_L[\sin^2(\pi x/a) + \sin^2(\pi y/a)]
$
via the relations~\cite{zwerger}
$
t_0 = (4 E_r/\sqrt{\pi}) (V_L/E_r)^{3/4} \exp(-2\sqrt{V_L/E_r})
$
and
$
U_0 = - \sqrt{8 \pi} a_F E_r (V_L/E_r)^{3/4}/a.
$
Here $a$ is half of the laser wavelength which corresponds to the lattice spacing, 
$V_L$ is the depth of the optical lattice potential, $E_r = \hbar^2 \pi^2/(2 m_0 a^2)$ is 
the recoil energy and $a_F$ is the two-body scattering length in vacuum. 
The experimental parameters $V_L$, $a$ and $a_F$ can be tuned by varying the
laser intensity, the laser wavelength and the externally applied magnetic field via 
using the Feshbach resonances, respectively, which makes optical lattices ideal 
systems to simulate the Fermi-Hubbard Hamiltonian.

For numerical purposes the superfluid order parameter is assumed to be 
real ($\Delta_i = \Delta_i^*$). This is sufficient to describe the nonmodulating 
and the spatially modulating superfluid phases in addition to the normal and the 
band insulator phases. We also take $U_0 = 3t_0$ and $V_0 = \alpha_0 a^2/2 = 0.02t_0$ 
as the strength of the weak onsite interactions and the weak trapping potentials, 
respectively, and perform calculations on a two-dimensional square lattice 
with a length of $L = 50a$ in both directions.
The trap center is located at $\mathbf{r_c} \equiv (x=0a, y=0a)$. 
We want to emphasize that similar calculations can be also performed 
for three-dimensional optical lattices. However they are computationally 
much more demanding and we do not expect any qualitative difference 
between our two-dimensional results and the three-dimensional ones.

We fix the total number of fermions $N = N_\uparrow + N_\downarrow$ to
$N \approx 270$ (corresponding to $\mu \approx 0t_0$) in the low density and 
to $N \approx 1570$ (corresponding to $\mu \approx 5t_0$) in the high density
case where $\mu = (\mu_\uparrow+\mu_\downarrow)/2$, while
we vary the population imbalance $P = (N_\uparrow - N_\downarrow)/N$ or equivalently 
$\delta \mu = (\mu_\uparrow - \mu_\downarrow)/2$.
For these parameters it is important to notice that the trapping potential
provides a soft boundary leading to a finite system, and therefore it simplifies 
the numerical calculations considerably in comparison to infinite systems. 
Next we present self-consistent solutions of Eq.~(\ref{eqn:bdg}) with the order 
parameter and the number equations.

\begin{figure} [htb]
\centerline{\scalebox{0.35}{\hskip 0mm \includegraphics{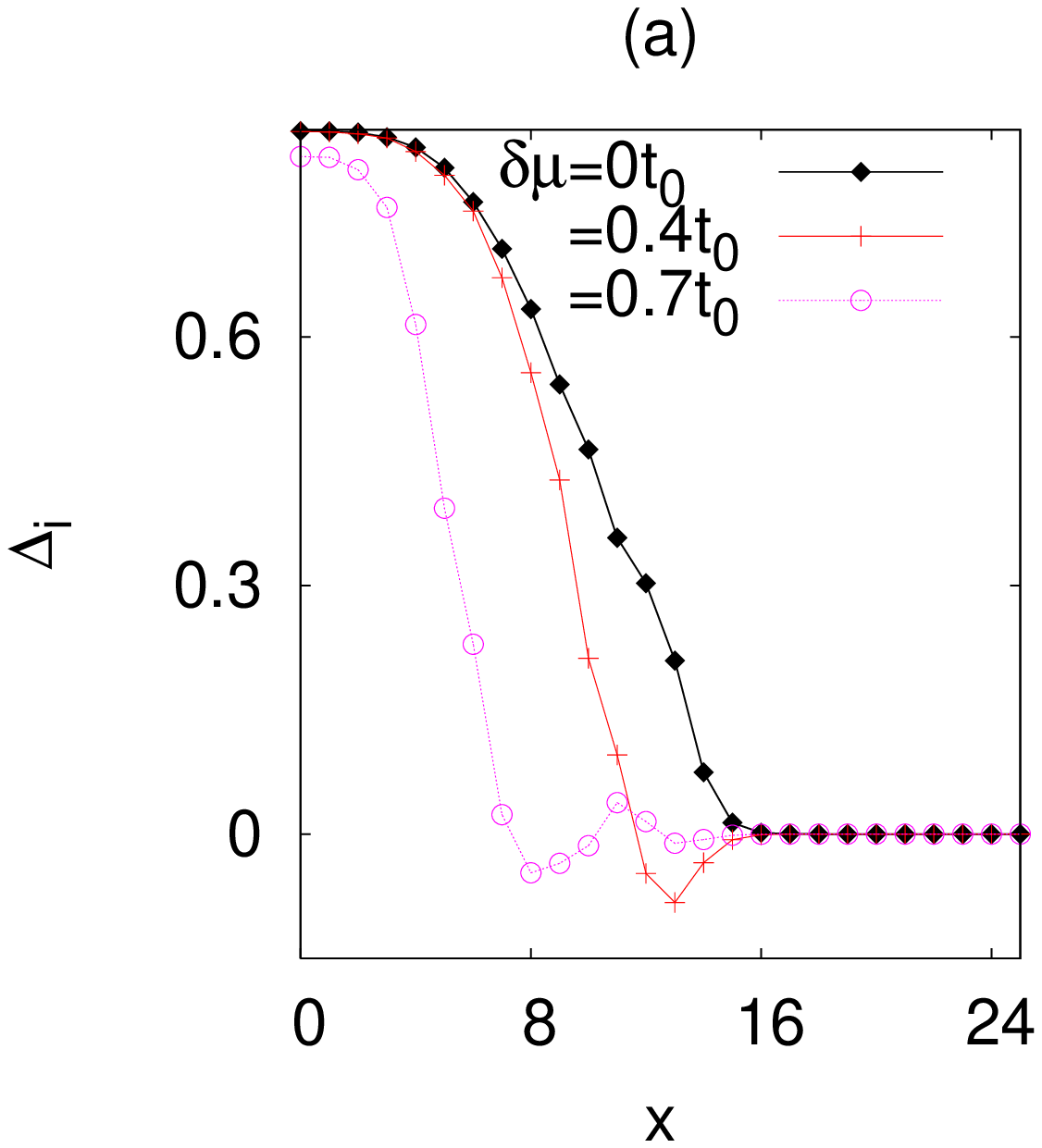} 
\hskip 0mm \includegraphics{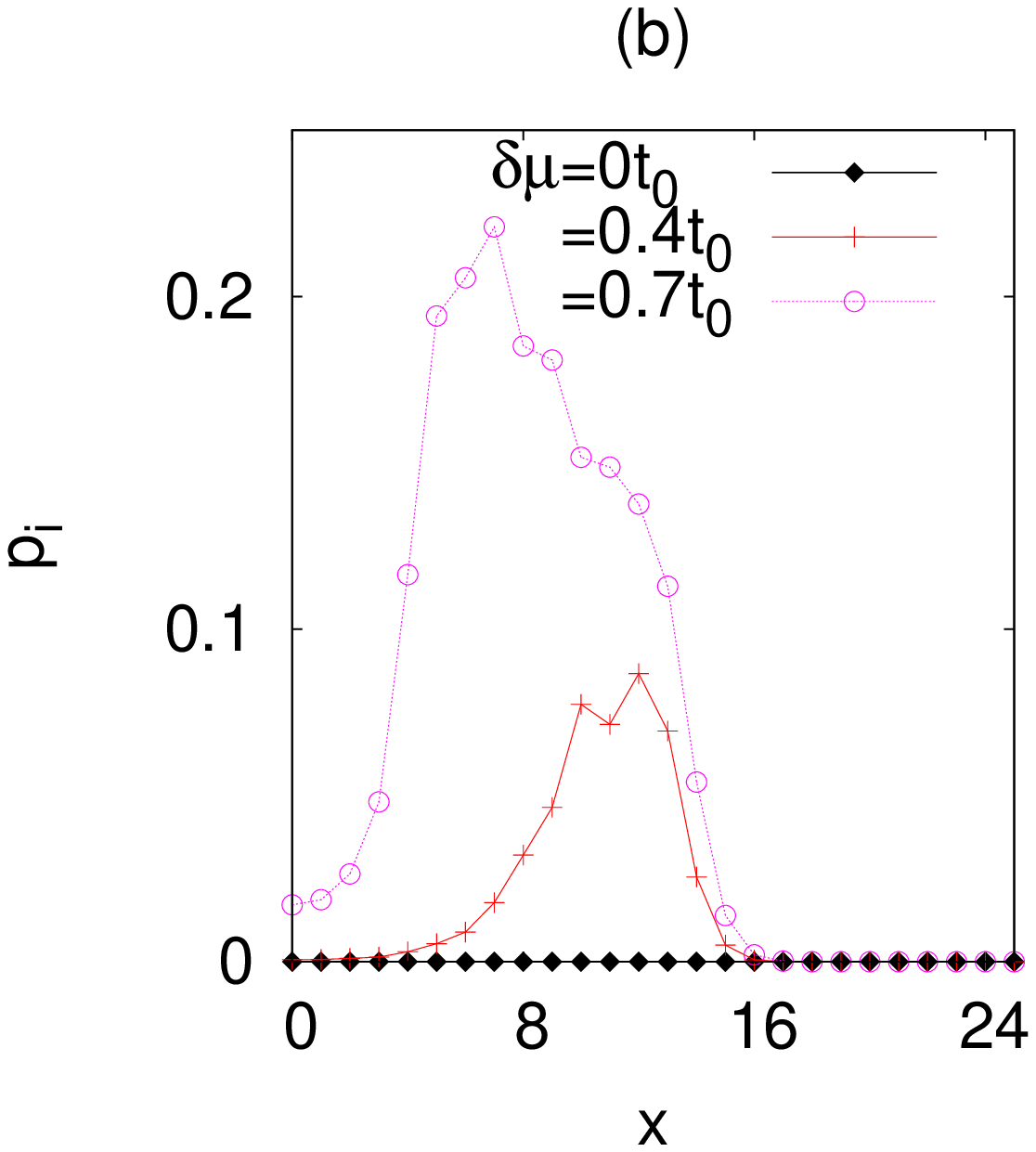}}}
\caption{\label{fig:low.2d} (Color online)
We show (a) the order parameter $\Delta_i$ (in units of $t_0$) and (b) the 
population difference $p_i = n_{i,\uparrow} - n_{i,\downarrow}$ (per lattice site)
for the low density case as a function of distance $x$ (in units of $a$) from 
the trap center. Here $y = 0a$.
}
\end{figure}
\textit{(I) Low density mixtures}:
In Fig.~\ref{fig:low.2d} we show the superfluid order parameter $\Delta_i$ and the 
population difference per lattice site $p_i = n_{i,\uparrow} - n_{i,\downarrow}$ for 
the low density case where $N \approx 270$. When $U_0 = 0$ and 
$N_\uparrow = N_\downarrow = N/2$, the maximum filling of this case corresponds 
to an almost half-filled band with $n_{i,\sigma} \approx 0.5$ at the trap center.
For such low densities we expect that our results for the trapped
mixtures with an optical lattice to recover the previously obtained results 
for the trapped dilute mixtures without an optical lattice~\cite{mizushima, torma, liu-mixture}.
This occurs when the interparticle separation becomes much longer than $a$ 
such that the particles do not feel the presence of a lattice potential. 
However to understand the ground state phases of population imbalanced mixtures, 
it is very illustrative to first discuss the population balanced case.

For a weakly attracting population balanced mixture with $U_0 = 3t_0$ and 
$\delta \mu = 0$, the order parameter $\Delta_i$ is finite around 
the trap center for distances $|\mathbf{r_i}| \lesssim 15a$, and therefore the 
ground state corresponds to a BCS type superfluid. For longer distances 
$|\mathbf{r_i}| \gtrsim 16a$  away from the trap center, $\Delta_i$ gradually 
decreases until it eventually vanishes when the densities become very 
low $n_{i,\uparrow} = n_{i,\downarrow} \approx 0$. These features can be 
seen in Fig.~\ref{fig:low.2d}(a), and they are in good agreement with the 
earlier experiments involving population balanced mixtures without an optical 
lattice~\cite{djin, rgrimm, jthomas, csalomon, rhulet}.

In the case of population imbalanced mixtures, we find that $\Delta_i$ 
modulates in the radial direction towards the trap edges to accommodate 
the unpaired fermions. However $\Delta_i$ decreases with increasing 
population imbalance as shown in Fig.~\ref{fig:low.2d}(a), and it vanishes 
entirely beyond a critical imbalance signaling a transition from the superfluid 
to the normal phase. These features can be seen in Fig.~\ref{fig:low.2d}(a) 
where $\delta \mu = 0.4t_0$ and $\delta \mu = 0.7t_0$ corresponding 
to $P \approx 0.12$ and $P \approx 0.34$, respectively. Similar spatial 
modulations have been recently found also in dilute population imbalanced 
mixtures without an optical lattice~\cite{mizushima, torma, liu-mixture}, 
however they have not yet been observed in the current 
experiments~\cite{mit, rice, mit-2, rice-2}. In contrast to our BdG results, 
the LDA type methods exclude the possibility of order parameter 
modulations and therefore they fail to produce such a spatially 
modulated superfluid phase which is one of the possible candidates 
for the ground state. 

In the recent theoretical works on dilute population imbalanced mixtures without an 
optical lattice, such spatial modulations have been suggested as 
signatures for the FFLO type superfluidity by some authors~\cite{mizushima, torma}
and as finite size effects by some others~\cite{liu-mixture}. Here we remind 
that the FFLO type superfluidity is characterized by the formation of Cooper 
pairs with nonzero center-of-mass momentum, in contrast with the BCS type 
superfluidity where Cooper pairs have zero center-of-mass momentum~\cite{casalbuoni}.
Therefore in two- and three-dimensional systems it is an open question whether 
these spatial modulations are related to the FFLO superfluidity or are simply 
finite size effects. However we also remind that the exact ground state 
phase diagram of one-dimensional systems have been recently 
calculated~\cite{1Dorso, 1Dhui, 1Dtezuka, 1Dbatrouni} showing that the 
superfluid phase has FFLO structure in trapped as well as infinite systems.

In Fig.~\ref{fig:low.2d}(b) we show that the unpaired fermions are pushed away 
from the trap center towards the trap edges and they have a maximum at the 
position where $\Delta_i$ changes sign. This is because spatially bound 
Andreev type states form around the nodes of $\Delta_i$, and the occupation 
of these bound states is different for $\uparrow$ and $\downarrow$ 
fermions~\cite{mizushima}. Since $\mu_\uparrow > \mu_\downarrow$ when
$N_\uparrow > N_\downarrow$, the $\uparrow$ fermions mostly occupy 
these states leading to the single peak structure. This feature is in good agreement 
with the recent experiments on dilute population imbalanced mixtures without 
an optical lattice~\cite{mit, rice, mit-2, rice-2}.
However in contrast with the trapped mixtures without an optical lattice,
both $\Delta_i$ and $p_i$ have $C_4$ symmetry which is consistent with 
the underlying symmetry of the square lattice. Here we notice that the LDA 
type methods always produce results with rotational symmetry and therefore
they are not strictly applicable to optical lattices.
Having shown that the ground state phases of low density mixtures in optical lattices 
are qualitatively similar to those of the dilute mixtures without an optical 
lattice, next we discuss the high density mixtures.

\begin{figure} [htb]
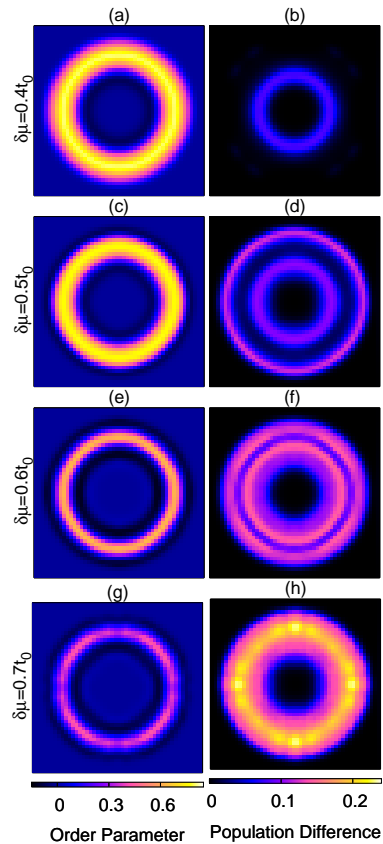

\centerline{\scalebox{0.45}{\includegraphics{gap-g3-mu5-h4.epsi} \includegraphics{mag-g3-mu5-h4.epsi}}}
\centerline{\scalebox{0.45}{\includegraphics{gap-g3-mu5-h5.epsi} \includegraphics{mag-g3-mu5-h5.epsi}}}
\centerline{\scalebox{0.45}{\includegraphics{gap-g3-mu5-h6.epsi} \includegraphics{mag-g3-mu5-h6.epsi}}}
\centerline{\scalebox{0.45}{\includegraphics{gap-g3-mu5-h7.epsi} \includegraphics{mag-g3-mu5-h7.epsi}}}
\caption{\label{fig:high.3d} (Color online)
We show the order parameter $\Delta_i$ (on the left, in units of $t_0$) and
population difference $p_i = n_{i,\uparrow} - n_{i,\downarrow}$ (on the right, per lattice site)
for the high density case on a two-dimensional square lattice with $50a \times 50a$ sites. 
Here the chemical potentials are such that 
(a,b) $\delta \mu = 0.4t_0$;
(c,d) $\delta \mu = 0.5t_0$;
(e,f) $\delta \mu = 0.6t_0$; and
(g,h) $\delta \mu = 0.7t_0$.
}
\end{figure}
\begin{figure} [htb]
\centerline{\scalebox{0.35}{\hskip 0mm \includegraphics{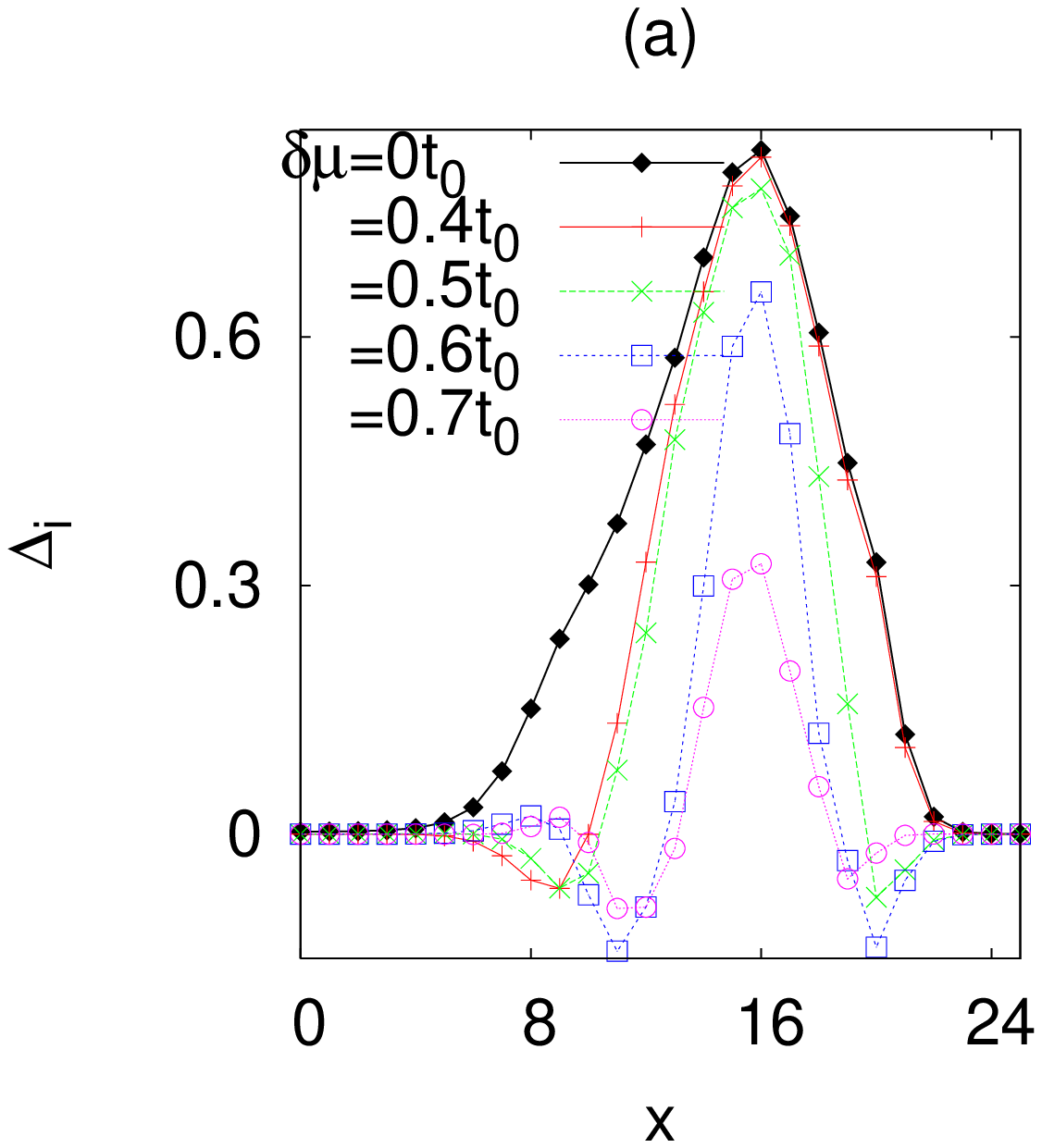} \hskip 0mm \includegraphics{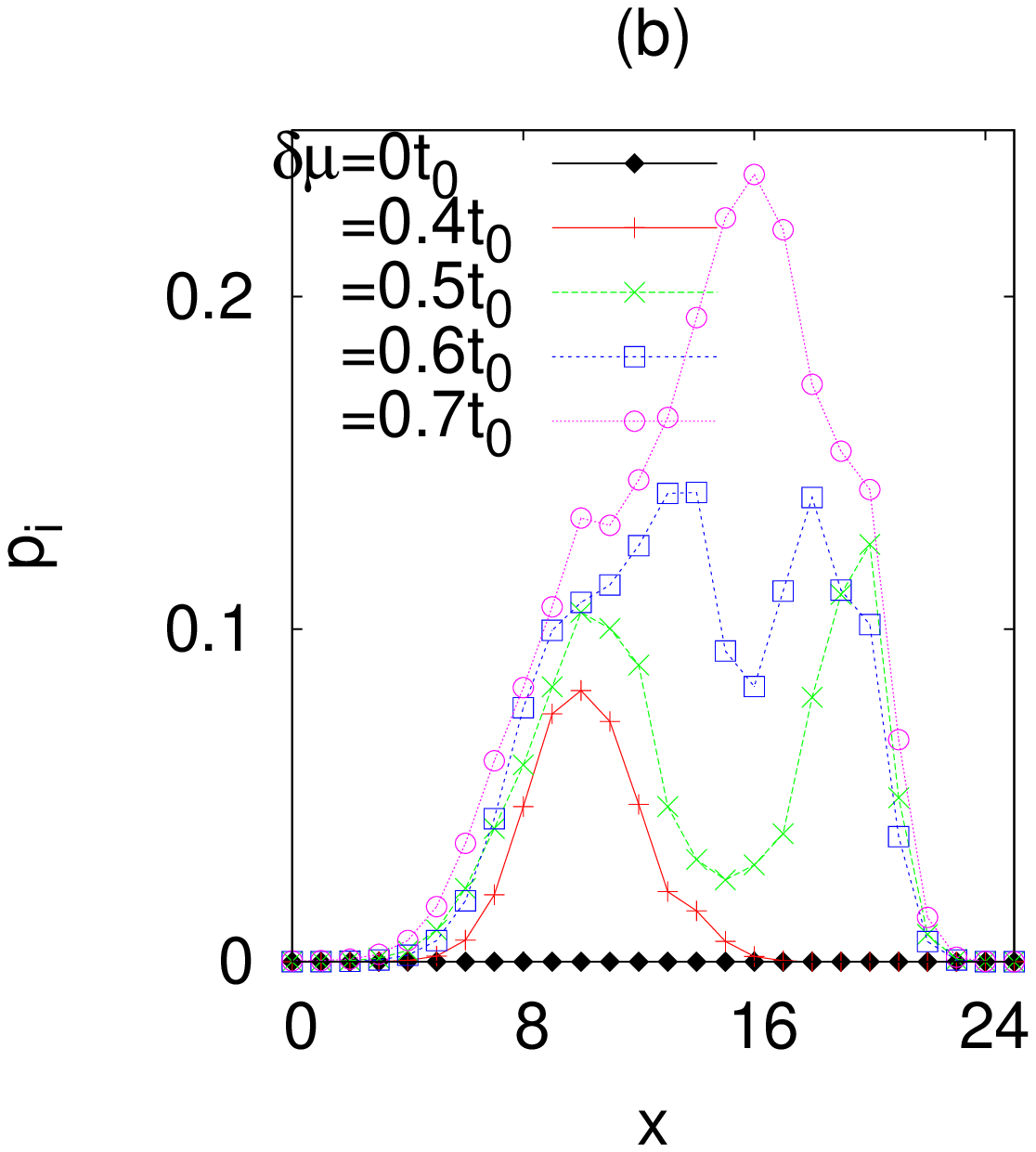}}}
\caption{\label{fig:high.2d} (Color online)
We show (a) the order parameter $\Delta_i$ (in units of $t_0$) and (b) the 
population difference $p_i = n_{i,\uparrow} - n_{i,\downarrow}$ (per lattice site)
for the high density case as a function of distance $x$ (in units of $a$) 
from the trap center. Here $y = 0a$.
}
\end{figure}

\textit{(II) High density mixtures}:
In Figs.~\ref{fig:high.3d} and~\ref{fig:high.2d} we show the superfluid order parameter 
$\Delta_i$ and the population difference per lattice site $p_i$ for the high density 
case where $N \approx 1570$. When $U_0 = 0$ and $N_\uparrow = N_\downarrow = N/2$,
the maximum filling of this case corresponds to a fully-filled band with 
$n_{i,\sigma} = 1$ near the trap center.
For such high densities the ground state phases are very different from those of the
low density systems as can be seen in Figs.~\ref{fig:high.3d}
and~\ref{fig:high.2d}. To understand these ground state phases of
population imbalanced mixtures, it is again very illustrative to first discuss 
the population balanced case.

For a weakly attracting population balanced mixture with $U_0 = 3t_0$ and 
$\delta \mu = 0$, we find that $\Delta_i = 0$ around the trap center for distances
$|\mathbf{r_i}| \lesssim 4a$. This signals the band insulator phase characterized 
by a fully-filled band where $n_{i,\uparrow} = n_{i,\downarrow}= 1$.
However since $n_{i,\uparrow} = n_{i,\downarrow} < 1$ away from the trap center,
$\Delta_i$ becomes finite signaling a transition from the band insulator to the superfluid 
phase. The maximum $\Delta_i$ occurs around $|\mathbf{r_i}| \approx 16a$ where
$n_{i,\uparrow} = n_{i,\downarrow}= 0.5$ corresponding to a half-filled band. 
This is purely a density of states ($D_i$) effect since
$\Delta_i \propto t_0 e^{-1/(U_0 D_i)}$ and $D_i$ has a maximum exactly at half-filling
due to particle-hole symmetry of the Fermi-Hubbard Hamiltonian.
For longer distances $|\mathbf{r_i}| \gtrsim 16a$ away from the trap center, $\Delta_i$ 
gradually decreases until it eventually vanishes for $|\mathbf{r_i}| \gtrsim 22a$
where $n_{i,\uparrow} = n_{i,\downarrow} \approx 0$.
These features can be seen in Fig.~\ref{fig:high.2d}(a) and they are 
very different from those of the low density case shown in Fig.~\ref{fig:low.2d}(a).
In contrast to our BdG results, the LDA type methods fail to describe the band 
insulator region with unit filling because the density profiles do not vary smoothly 
as a function of $|\mathbf{r_i}|$.

In the case of population imbalanced mixtures, we find that $\Delta_i$ 
modulates in the radial direction towards the trap center for low imbalance 
as shown in Fig.~\ref{fig:high.3d}(a) because $\Delta_i$ is a more slowly decreasing 
function of $|\mathbf{r_i}|$ towards the trap center than towards the trap edges 
when $\delta \mu = 0$. However $\Delta_i$ also modulates towards the trap edges 
with increasing imbalance as shown in Figs.~\ref{fig:high.3d}(c),~\ref{fig:high.3d}(d) 
and~\ref{fig:high.3d}(e). Characteristic features of these spatial modulations
are similar to those of the low density systems and they can be seen in Fig.~\ref{fig:high.2d}(a) 
where $\delta \mu = 0.4t_0$, $\delta \mu = 0.5t_0$, $\delta \mu = 0.6t_0$ and 
$\delta \mu = 0.7t_0$ corresponding to $P \approx 0.017$, $P \approx 0.058$, 
$P \approx 0.090$ and $P \approx 0.12$, respectively. 
Therefore high density mixtures in trapped optical lattices are also good candidates 
for observation of such exotic superfluid modulations. Further increasing the 
population imbalance gradually decreases $\Delta_i$ as shown in Fig.~\ref{fig:high.2d}(a),
until it vanishes entirely beyond a critical imbalance signaling a transition from
the superfluid to the normal phase.

In Figs.~\ref{fig:high.3d}(b) and~\ref{fig:high.2d}(b) we show  for low imbalanced 
mixtures that the density of unpaired fermions has a single peak at the position 
where $\Delta_i$ changes sign. However since $\Delta_i$ also modulates 
towards the trap edges for intermediate imbalance, the unpaired fermions have 
double peaks in their density as shown in Figs.~\ref{fig:high.3d}(d),
~\ref{fig:high.3d}(f) and~\ref{fig:high.2d}(b). Furthermore since $\Delta_i$ vanishes 
with further increase in imbalance, these two peaks merge leading to a single peak 
which is shown in Figs.~\ref{fig:high.3d}(h) and~\ref{fig:high.2d}(b). Notice that
similar to the low density case both $\Delta_i$ and $p_i$ have $C_4$ 
symmetry which is consistent with the underlying symmetry of the square lattice.

\textit{Conclusions}:
To conclude we used the BdG method to analyze the ground state phases of 
population imbalanced fermion mixtures in harmonically trapped optical 
lattices. First we showed that the phase structure of low density mixtures 
in optical lattices are qualitatively similar to those of the dilute mixtures 
without an optical lattice. Then we discussed high density mixtures and 
found qualitatively different results. In both cases we found that the 
superfluid order parameter modulates spatially but it is an open 
question whether these modulations are related to the FFLO superfluidity 
or are simply finite size effects. Lastly we compared our BdG results with 
the LDA ones and argued that the LDA type methods are not sufficient to describe 
especially the high density mixtures in harmonically trapped optical lattices.

\end{document}